\documentclass[twocolumn,showpacs,preprintnumbers,floatfix]{revtex4-1}
\usepackage[english]{babel}
\usepackage[latin1]{inputenc}
\usepackage[T1]{fontenc}
\usepackage{lmodern}
\usepackage{graphicx}                   
\usepackage{braket}
\usepackage{textcomp}
\usepackage{amsmath}
\usepackage{amssymb}
\usepackage{braket}
\usepackage{color}

\newcommand\leqt[1]{\protect\label{eq:#1}}
\newcommand\reqtn[1]{\ref{eq:#1}}
\newcommand\reqt[1]{(\reqtn{#1})}
\renewcommand{\d}[1]{\mathrm{d}{#1}}

\graphicspath{{pictExper/}{pict/}{}}
\newcommand\pictc[5]{\begin{figure}[hl]
                      \centerline{
\includegraphics[width=#1\columnwidth,height=0.7\textheight,keepaspectratio]{#3}}
                      \protect\caption{\protect\label{fig:#4} #5}
                   \end{figure}            }
\newcommand\pict[4][1]{\pictc{#1}{!tb}{#2}{#3}{#4}}
\newcommand\rpict[1]{\ref{fig:#1}}

\begin{document}

\title{Biphoton generation in quadratic waveguide arrays: A classical optical simulation}
\normalsize

\author{M. Gr{\"a}fe$^{1,\ast}$, A. S. Solntsev$^{2,\ast}$, R.~Keil$^1$, A.~A.~Sukhorukov$^2$, M.~Heinrich$^1$, A.~T{\"u}nnermann$^1$, S.~Nolte$^1$, A.~Szameit$^1$,
and Yu.~S.~Kivshar$^2$}

\affiliation{$^1$Institute of Applied Physics, Abbe Center of Photonics, Friedrich-Schiller-Universit\"{a}t, Max-Wien-Platz 1, 07743 Jena, Germany}

\affiliation{$^2$Nonlinear Physics Centre and Centre for Ultra-high Bandwidth Devices for Optical Systems (CUDOS), Research School of Physics and
Engineering, Australian National University, Canberra, ACT 0200, Australia}

\affiliation{$^{\ast}$ both authors contributed equally}

\bigskip
\begin{abstract}
Quantum entanglement, the non-separability of a multipartite wave function, became essential in understanding the non-locality of quantum mechanics. In optics, this non-locality can be demonstrated on impressively large length scales, as photons travel with the speed of light and interact only weakly with their environment. With the discovery of spontaneous parametric down-conversion (SPDC) in nonlinear crystals, an efficient source for entangled photon pairs, so-called biphotons, became available. It has recently been shown that SPDC can also be implemented in nonlinear arrays of evanescently coupled waveguides which allows the generation and the investigation of correlated quantum walks of such biphotons in an integrated device.
Here, we analytically and experimentally demonstrate that the biphoton degrees of freedom are entailed in an additional spatial dimension, therefore the SPDC and the subsequent quantum random walk in one-dimensional (1D) arrays can be simulated through classical optical beam propagation in a
two-dimensional (2D) photonic lattice. Thereby, the output intensity images directly represent the biphoton correlations and exhibit a clear violation of a Bell-type inequality.
\end{abstract}


\maketitle

As proven by Bell \cite{Bell}, nonlocal quantum correlations play the central role in the understanding of the famous Einstein-Podolsky-Rosen
gedankenexperiment \cite{Einstein:1935-777:PREV}. In the realm of optics, the most prominent source of such non-local correlations are entangled
photon pairs, so-called biphotons \cite{Shih:2011:IntroductionQuantum}. They give rise to various applications such as quantum cryptography
\cite{Ekert:1992-1293:PRL}, teleportation~\cite{Bouwmeester:1997-575:NAT, *Boschi:1998-1121:PRL, *Pan:1998-3891:PRL, *Furusawa:1998-706:SCI}, and
quantum computation~\cite{Bouwmeester:2000:QuantumInformation}. A particular robust approach to realize strong quantum correlations of path-entangled
photons in compact settings is their propagation in optical waveguide arrays~\cite{Bromberg:2009-253904:PRL, Peruzzo:2010-1500:SCI} that provide a
unique tool for the experimental analysis of a spatially discrete, continuous-time, quantum walks with highly controllable parameters. However, in
these settings the photons are typically generated \textit{before} they are launched into the array, usually via spontaneous parametric
down-conversion (SPDC) in bulk optical components \cite{Shih:2011:IntroductionQuantum}. Recently, it was suggested theoretically
\cite{Solntsev:2012-23601:PRL} that biphotons can be generated directly in a quadratically nonlinear waveguide array through SPDC, and quantum walks
of the generated photons can give rise to non-classical correlations at the output of this monolithic integrated optical device.

Remarkably, various quantum phenomena can be simulated via purely classical light propagation in waveguide arrays, serving as an optical test-bed of
fundamental physical effects without the need of an intricate single photon setup \cite{Keil:2011-103601:PRL, *Longhi:2009oq, *Szameit:2010nx}. In particular, the non-classical 1D evolution of correlated quantum particles can be mapped onto a classical 2D evolution of a single wave packet, as was predicted recently by Longhi~\cite{Longhi:2008-193902:PRL, Longhi:2011-3248:OL}. By increasing the dimensionality of the structure, the same dynamics is obtained as if the number of participating photon wavepackets were effectively doubled. This concept was recently applied in the context of a discrete-time random walk of coherent light~\cite{Schreiber:2012dq}. In this article we will show that linear propagation of classical
light beams can be used to simulate the nonlinear effect of SPDC leading to photon pair generation and quantum walks of the generated biphotons.
Specifically, we prove theoretically and demonstrate experimentally that the quantum correlations of biphotons generated in a quadratically nonlinear
1D waveguide array with pump waveguides at the edges 
can be mapped onto the linear propagation in a specially designed 2D system.
This analogy enables us to simulate, with classical light, the quantum coincidence counts for biphotons and the breaking of a Bell-type inequality.

\section{Results}
We first extend the theory of photon-pair generation in quadratic waveguide arrays from Ref.~\cite{Solntsev:2012-23601:PRL} to finite waveguide arrays. This is especially important for an experimental realization, where the number of waveguides is naturally limited.
We consider a type-I SPDC process with continuous-wave or narrow-band pump of frequency $\omega_p$, which converts pump photons into pairs of signal and idler photons with frequencies $\omega_{s,i}$. We assume spectral
filtering at half of the pump frequency, such that $\omega_{s,i} = \omega_p/ 2$. The signal and idler photons can tunnel between the neighboring
waveguides at a rate characterized by the coefficients $C_{s,i} = C$, whereas coupling of the pump can be neglected, due to the strong dispersion of the evanescent coupling~\cite{Lederer:2008-1:PRP, Bromberg:2009-253904:PRL,
Peruzzo:2010-1500:SCI, Solntsev:2012-23601:PRL, Szameit:2007cr}.
 Taking this into account and following Refs.~\cite{Christ:2009-33829:PRA, Grice:1997-1627:PRA,
DiGiuseppe:2002-13801:PRA} one can derive the differential equations for the biphoton wave function $\Psi$ (see Methods)
\begin{equation} \leqt{real_space_maintext}
    \begin{split}
        & i \frac{\d{\Psi_{n_s, n_i}(z)}}{\d{z}} =
        - C \left[\Psi_{n_s+1, n_i} + \Psi_{n_s-1, n_i} + \Psi_{n_s, n_i+1} \right. \\
        &\left.  + \Psi_{n_s, n_i-1} \right]
        + i \sum_{n_p} d_{\text{eff}} A_{n_p} \delta_{n_s,n_p} \delta_{n_i,n_p} \exp( i \Delta\beta^{(0)} z) ,
    \end{split}
\end{equation}
whereby $n_{s,i}$ label the waveguide index for signal and idler, respectively, $\delta$ denotes the Kronecker delta function and $\Delta\beta^{(0)}$ is the phase mismatch in a single waveguide. The pump field is distributed over the waveguides according to $A_{n}$ whereas $d_{\text{eff}}$ is the effective nonlinear coefficient. This equation governs the evolution of a biphoton wave function in a quadratic nonlinear waveguide array. The term in square brackets on the right-hand side describes the quantum walks, i.~e. the biphoton tunneling~\cite{Solntsev:2012-23601:PRL,Peruzzo:2010-1500:SCI, Solntsev:2012-23601:PRL} between the waveguides, and corresponds to the model derived in Ref.~\cite{Longhi:2011-3248:OL} in case of a linear waveguide array. In addition, our model also describes the nonlinear process of biphoton generation through SPDC, which is introduced by the last term. Note that both photons are always created at the same spatial location, which is expressed mathematically through $\delta$-functions, but the biphotons subsequently spread out through correlated quantum walks.

After establishing the key model Eq.~\reqt{real_space_maintext}, we now demonstrate how the biphoton wave function can be simulated classically.
Note that
Eq.~\reqt{real_space_maintext} is formally equivalent to the coupled-mode equations for classical beam propagation in a square linear 2D waveguide array~\cite{Lederer:2008-1:PRP, Longhi:2011-3248:OL} with additional source terms describing the pump. In the following, we demonstrate that for a pump exclusively coupled to the outermost waveguides of the 1D array, the effect of these terms can be simulated classically by introducing additional waveguides close to the corners of an equivalent 2D array, as shown in Fig.~\rpict{sketchStructures}. We consider the case when the additional waveguide modes only weakly overlap with the modes at the array corners, and the corresponding coupling coefficient $C_c$ is very small compared to the array, $C_c \ll C$. If the light is launched only into the additional waveguides, then in the undepleted pump approximation the classical field evolution in those waveguides is $\Psi_{\text{left}}(z) = \Psi_{\text{left}}(0) \exp(i \beta_{\text{left}} z)$ and
$\Psi_{\text{right}} =  \Psi_{\text{right}}(0) \exp(i \beta_{\text{right}} z)$, where $\Psi_{\text{left}}(0)$ and $\Psi_{\text{right}}(0)$ are the input amplitudes, $\beta_{\text{left}}$ and
$\beta_{\text{right}}$ are the mode detunings of the additional waveguides with respect to a single waveguide in the array. Here, we consider the case where
both additional waveguides are identical to the guides of the lattice ($\beta_{\text{left}} = \beta_{\text{right}} = 0$). Then, the classical field
evolution in the array follows Eq.~\reqt{real_space_maintext} with the effective pump amplitudes in the edge waveguides $A_1 = C_c \Psi_{\text{left}}(0)$, $A_N = C_c
\Psi_{\text{right}}(0)$ (there is zero effective pump, $A_n=0$, for $1<n<N$), and the effective detuning $\Delta\beta^{(0)} = 0$. This corresponds
to perfectly phase-matched SPDC in the 1D-array, which has been shown to produce the most pronounced quantum correlations \cite{Solntsev:2012-23601:PRL}.
\pict{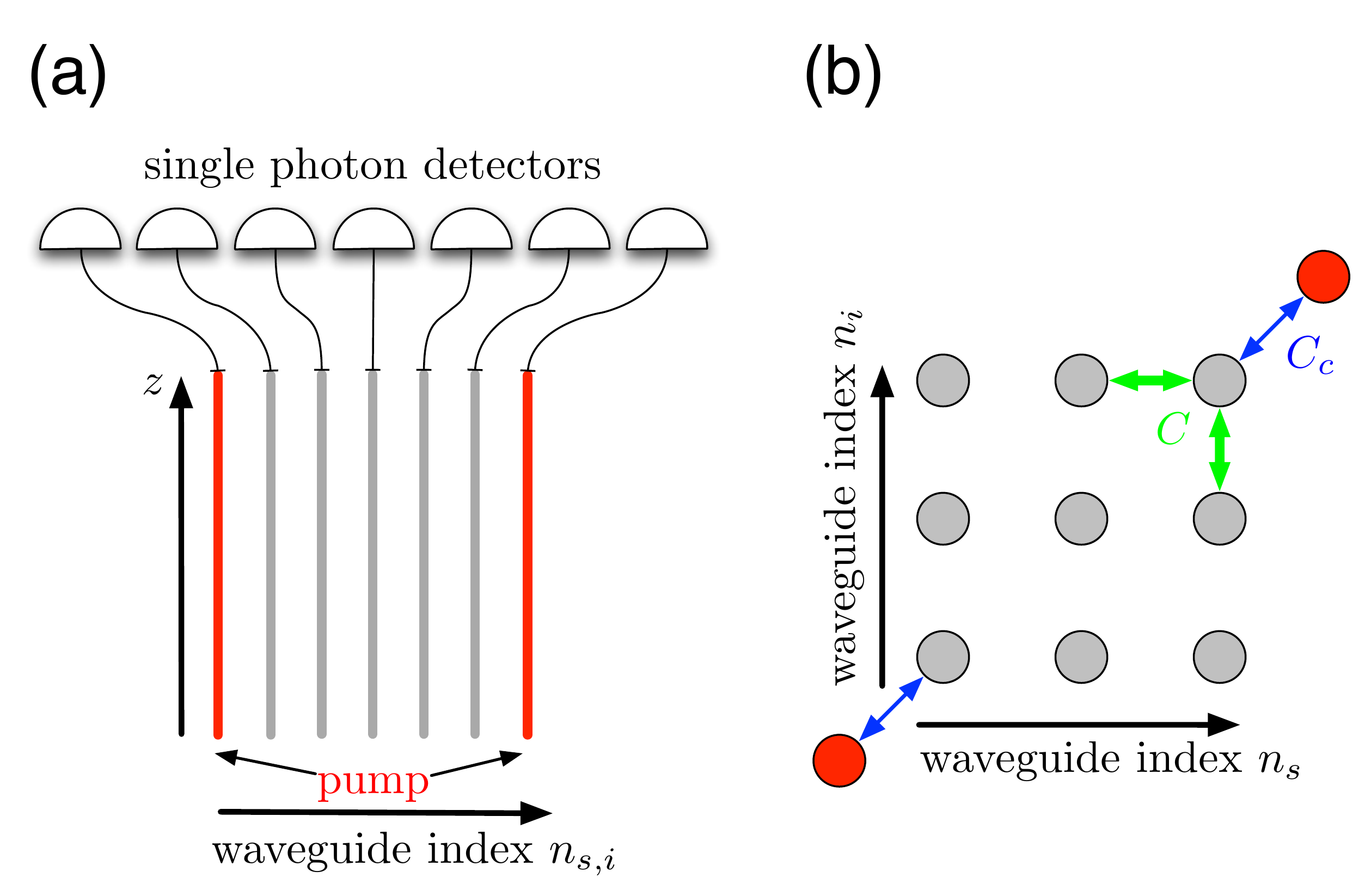}{sketchStructures}{\textbf{Settings} \textbf{(a)}~Sketch of a 1D quadratic nonlinear waveguide array containing $N=7$ waveguides with two
pumps coupled to the edge waveguides leading to biphoton generation via SPDC, and the output photon detectors. \textbf{(b)}~Cross-section of a 2D
waveguide array for the classical optical simulation of biphoton generation via SPDC consisting of $3\times3$ waveguides. The additional waveguides
represent the pump beam coupled to the edge waveguides.}
Based on the established mathematical equivalence classical light propagation in the linear array structure shown in Fig.~\rpict{sketchStructures}(b)
can be employed to simulate the quantum properties of biphotons generated in a quadratic nonlinear array shown in Fig.~\rpict{sketchStructures}(a).
In particular the photon number correlation $\Gamma_{n_s,n_i} =
\braket{\hat{a}^{\dagger}_{n_s}\hat{a}^{\dagger}_{n_i}\hat{a}_{n_i}\hat{a}_{n_s}}_{\Psi}$, which describes the probability of simultaneous detection
of photons at the output of the waveguides with number $n_s$ and $n_i$, is of interest. Thereby, $\hat{a}_{n_{s(i)}}$ denotes the photon annihilation
operator in these waveguides. In the classical simulation the output intensity distribution $I_{n_s,n_i} = |\Psi_{n_s,n_i}|^2$ corresponds directly
to $\Gamma$:
\begin{equation} \leqt{Gamma}
   \Gamma_{n_s,n_i}^{(\text{cl})}= I_{n_s,n_i} \left ( \sum_{n_s,n_i} I_{n_s,n_i} \right )^{-1}\widehat{=} \;\Gamma_{n_s,n_i} \; .
\end{equation}
%
\pict{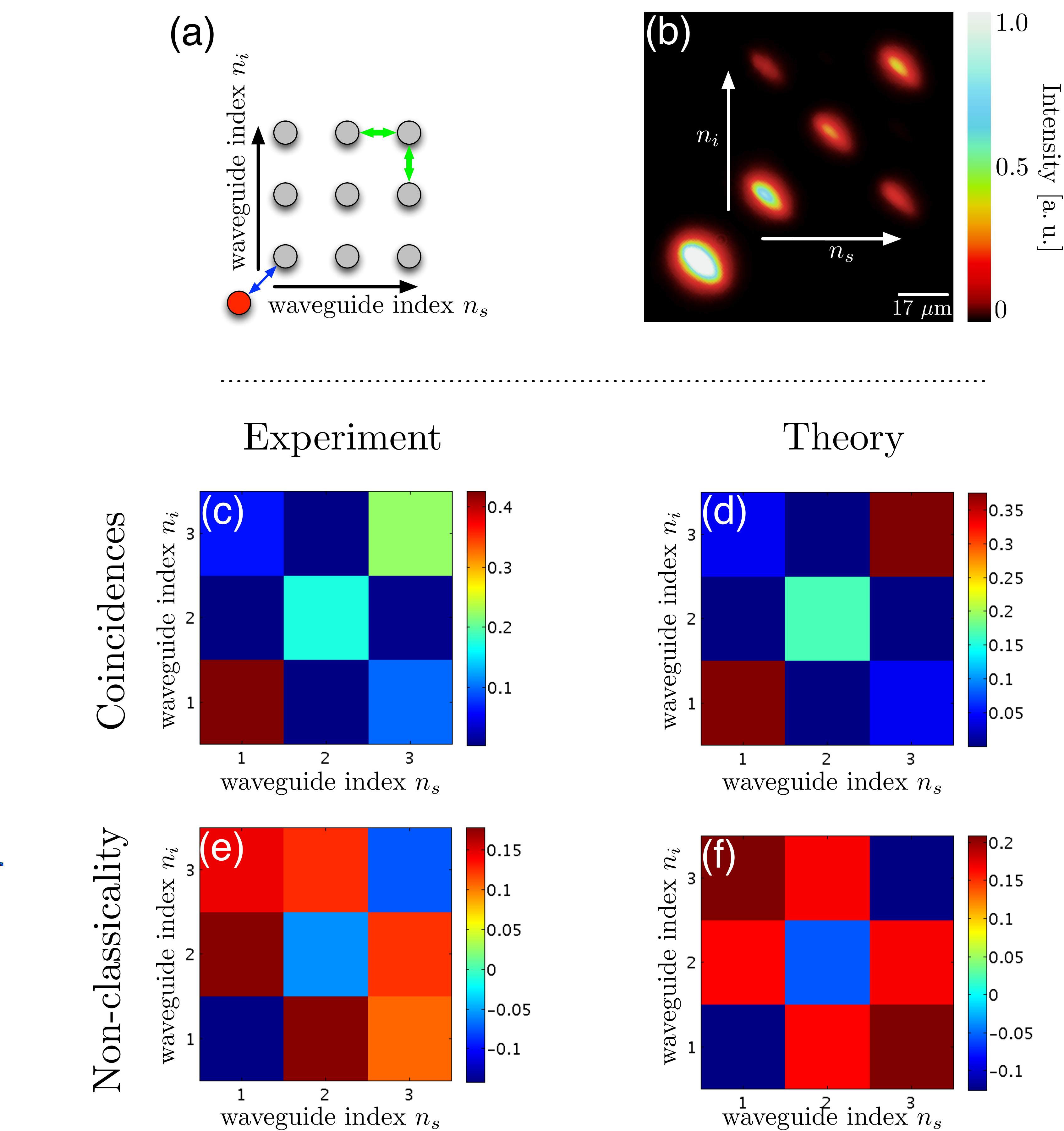}{fig1pump}{\textbf{Results for one pump waveguide} \textbf{(a)}~Sketch of the linear 2D waveguide array simulating quantum correlations of
biphotons generated by SPDC with one pump at the edge waveguide. \textbf{(b)}~CCD camera image of the output light distribution when only the pump
waveguide is excited. \textbf{(c,d)}~Correlation map of the simulated 1D quantum system: \textbf{(c)}~Extracted from intensity measurement shown
in~\textbf{(b)} and \textbf{(d)}~calculated numerically based on Eq.~\reqt{real_space_maintext}. \textbf{(e,h)}~Simulated non-classicality function
determined with Eq.~\reqt{nonclassicality} for \textbf{(e)}~experimental and \textbf{(f)}~numerical correlations from~\textbf{(c)} and~\textbf{(d)},
respectively. }
\pict{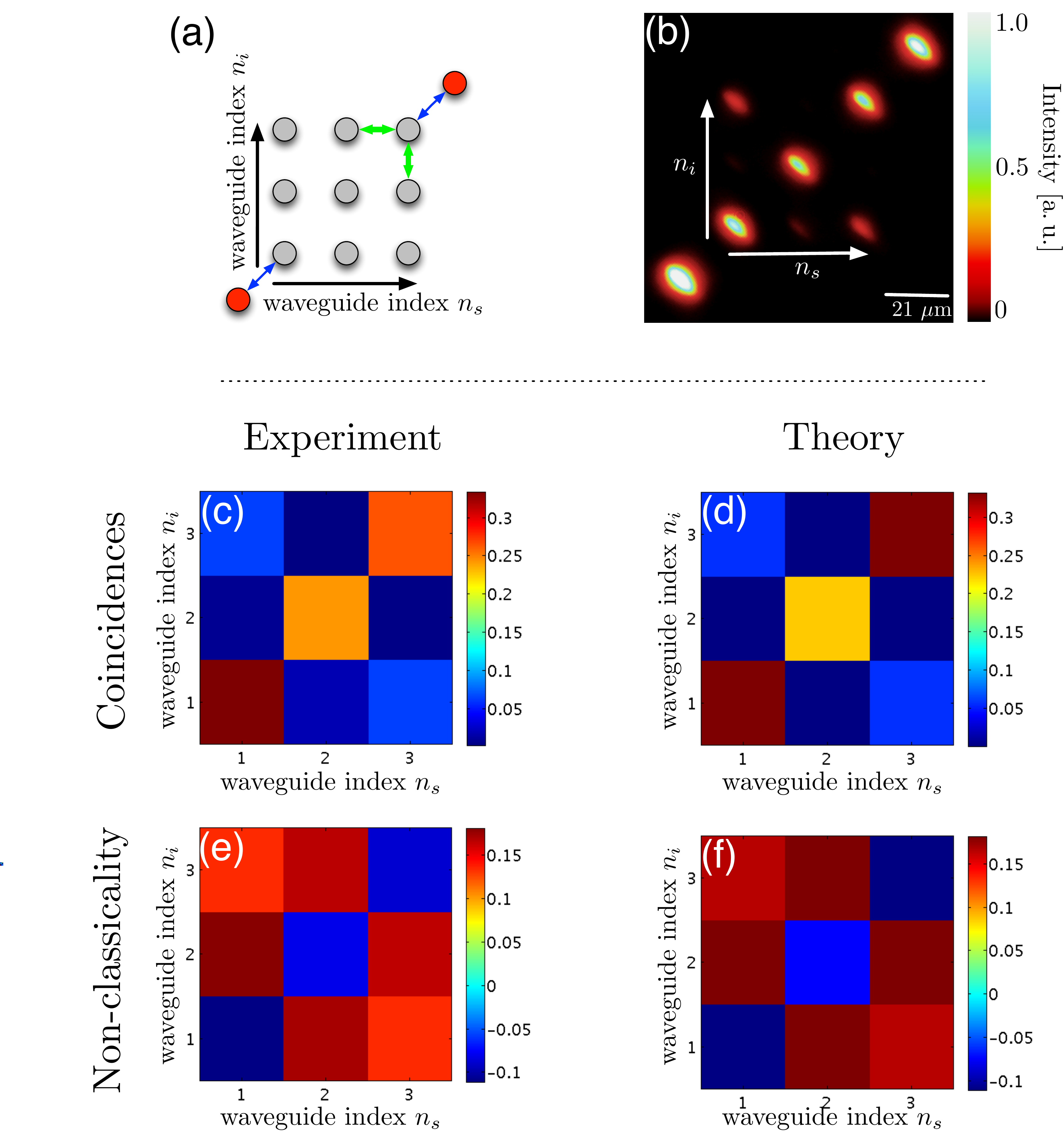}{fig2pump}{\textbf{Results for two pump waveguides} \textbf{(a)}~Sketch of the linear 2D waveguide array simulating quantum correlations
of biphotons generated by SPDC with equal pumps at the two edge waveguides. \textbf{(b-f)}~Experimental and theoretical results, notations are the
same as in Fig.~\rpict{fig1pump}. }
To verify the correct operation of our simulator, we first implemented a system with a single outer pump waveguide coupled to the corner of  $3 \times 3$ waveguide array
[Fig.~\rpict{fig1pump}(a)]. For the investigation light is launched into the device and imaged onto a CCD at the output (see Methods).
The experimental output intensity distribution is shown in Fig.~\rpict{fig1pump}(b). We perform a simple scaling of the measured intensity according to Eq.~\reqt{Gamma} and determine the two-photon correlation function as shown in Fig.~\rpict{fig1pump}(c). These results closely match the theoretical predictions presented in Fig.~\rpict{fig1pump}(d).

After verifying the correct operation in the single-pump regime, we return to the initially considered case of two pump waveguides at the corners of the 2D lattice as shown in Fig.~\rpict{fig2pump}(a). The challenge here was to have light of the same amplitude and phase in both pump waveguides. The implementation is discussed in Methods. We likewise use the output intensity distribution [Fig.~\rpict{fig2pump}(b)] to calculate the photon number correlation [Fig.~\rpict{fig2pump}(c)] which agrees very well with the simulation shown in Fig.~\rpict{fig2pump}(d).
\section{Discussion}
After obtaining the correlation functions from the output intensities, we determine in accordance with Ref.~\cite{Peruzzo:2010-1500:SCI}, the similarity $S=\left(\sum_{i,j}\sqrt{\Gamma_{i,j}^{\text{num}}\,\Gamma_{i,j}^{\text{exp}}}\right)^{2} / \left(\sum_{i,j}\Gamma_{i,j}^{\text{num}}\,\sum_{i,j}\Gamma_{i,j}^{\text{exp}}\right)$ between the numerically calculated ($\Gamma_{i,j}^{\text{num}}$) and experimentally obtained ($\Gamma_{i,j}^{\text{exp}}$) distributions as $S = 0.954$ for the case of one pump waveguide and $S = 0.972$ for two pump waveguides.

In a further step we demonstrate that our classical optical simulator can model non-classical biphoton statistics. Following
Ref.~\cite{Peruzzo:2010-1500:SCI}, we calculate the \emph{non-classicality}
\begin{equation} \leqt{nonclassicality}
    V_{n_s,n_i} = \frac{2}{3}\sqrt{\Gamma_{n_s,n_s}\Gamma_{n_i,n_i}} - \Gamma_{n_s,n_i}\; .
\end{equation}
Values of $V_{n_s, n_i} > 0$, indicate true quantum behavior corresponding to a violation of Bell's inequality, which cannot occur in a purely classical setting. In the single pump case we plot the function $V_{n_s, n_i}$ for our experimental data in Fig.~\rpict{fig1pump}(e), which shows the presence of positive values in agreement with numerical simulations [Fig.~\rpict{fig1pump}(f)].
This proves that our 2D classical setting successfully simulates actual non-classical 1D statistics. Fig.~\rpict{fig2pump}(e) and Fig.~\rpict{fig2pump}(f) illustrate the experimentally and numerically calculated non-classicality. Also for this setting it is evident, that the nonclassical features can be simulated with high quality.

As evident from the strong overexposure of the CCD at the pump waveguide in Fig.~\rpict{fig1pump}(b) and fact that the majority of the light remains in the pump waveguides in Fig.~\rpict{fig2pump}(b), the undepleted pump approximation holds.

For both a single and double pump waveguides, the output correlation should be symmetric, $\Gamma_{n_s,n_i} = \Gamma_{n_i,n_s}$, due to the indistinguishability of two photons, and we see that this feature is properly reproduced through the 2D experimental intensity distributions where $I_{n_s,n_i} \simeq I_{n_i,n_s}$, see Figs.~\rpict{fig1pump}(b) and~\rpict{fig2pump}(b). For a double pump configuration, there appears additional symmetry due to identical pump amplitudes at the two boundaries, and we have in theory $I_{n_s,n_i} = I_{N + 1 - n_i,N + 1 - n_s}$ [Figs.~\rpict{fig2pump}(d)], whereas such symmetry is clearly broken for a single pump case [Figs.~\rpict{fig2pump}(d)]. In experiment we have only a slight asymmetry of the output intensity distributions due to imperfections of the three-waveguide input-coupler (see Methods). Nevertheless, for all cases we have similarity $S$ close to unity, as well as a good agreement of the experimental and simulated non-classicality which proves excellent accuracy of the classical optical simulator.




In conclusion, we have derived a model describing the evolution of a biphoton wave function in 1D quadratic nonlinear waveguide arrays, where the photons are generated through SPDC and undergo correlated quantum walks. We further demonstrated analytically as well as experimentally that the quantum biphoton dynamics can be simulated by a classical wave evolution in a linear 2D waveguide array with additional waveguides representing the pump. Hereby, the additional spatial dimension provides the degrees of freedom which are otherwise encoded in the two-particle dynamics. The classical measurements of the output light intensity directly simulate the quantum biphoton correlation function, in particular including the
regime of the breaking of a 1D Bell-like inequality.
\section{Methods}
\subsection{Derivation of the biphoton wave function}
In the following a detailed derivation of the differential equation for the biphoton wave function is presented. We start from the coupled mode
equations for the classical light amplitudes in waveguide $n$ ($1 \le n \le N$)
%
\begin{equation} \leqt{E1d}
    i \frac{d E_n}{ d z} + C(E_{n-1}+E_{n+1}) = 0 ,
\end{equation}
with the boundary conditions $E_0 \equiv E_{N+1} \equiv 0$. The eigenstates of the system are the supermodes $E_n^{(m)}(z) = \varepsilon_n^{(m)}
\exp( i \beta_m z)$, where $m$ is the mode number ($1 \le m \le N$), $\beta_m = 2 C \cos(k_m)$ is the propagation constant, $\varepsilon_n^{(m)} =
\sqrt{2/(N+1)} \sin( n k_m)$ is the amplitude distribution, and $k_m = \pi m / (N+1)$.
For the pump, the corresponding coupling coefficient $C_p$ would generally have a much smaller value compared to the signal and idler waves, $C_p \ll
C$, due to the weaker mode overlap between neighboring waveguides at higher frequencies~\cite{Lederer:2008-1:PRP}. Therefore, we neglect coupling
effects for the pump beam ($C_p \approx 0$), and assume its amplitude $A_n$ to be constant along the propagation.

Subsequently, we employ the mathematical approach of Refs.~\cite{Christ:2009-33829:PRA, Grice:1997-1627:PRA,
DiGiuseppe:2002-13801:PRA} developed for multi-mode waveguides,
and obtain the following expression for the two photon state $\ket{\Xi}$:
\begin{equation} \leqt{middle_state}
        |{ \Xi}(z) \rangle = B \sum_{m_i=1}^N \sum_{m_s=1}^N \!
           \Psi_{m_s, m_i}(z)
                   \hat{a}^{\dagger}(m_s) \,
                \hat{a}^{\dagger}(m_i) \,
                | 0 , 0 \rangle ,
\end{equation}
where
\begin{equation} \leqt{Psi}
    \begin{split}
           \Psi_{m_s, m_i}(z)  = & \exp(i \beta_{m_s} z + i \beta_{m_i} z) \\
           & \int_{0}^z \d{z'}
        \gamma_{m_s,m_i}
        \exp (i \Delta\beta_{m_s, m_i} z').
    \end{split}
\end{equation}
Here $B$ is a constant, $\hat{a}^{\dagger}$ are photon creation operators at the numbered supermode states, $| 0 , 0 \rangle$ is the vacuum state, $\Psi_{m_s, m_i}$ is a two-photon wave function, $\Delta \beta_{m_s, m_i} = \Delta\beta^{(0)} - \beta_{m_s} - \beta_{m_i}$ is a phase mismatch between the modes, and $\Delta\beta^{(0)}$ is the mismatch in a single waveguide. The value $\gamma_{m_s,m_i} = \sum_{n=1}^N d_{\text{eff}} A_n\varepsilon_n^{(m_s)} \varepsilon_n^{(m_i)}$ is the normalized spatial overlap of the
signal and idler distributions with the pump distribution.
We note that an integration over frequency is omitted, since we consider a frequency window narrower then the phase-matching bandwidth of the SPDC.

For our analysis, it is convenient to rewrite Eq.~\reqt{Psi} in the form of differential equations,
\begin{equation} \leqt{dPsi_m}
    \begin{split}
        i \frac{\d{\Psi_{m_s, m_i}(z)}}{\d{z}} =& - (\beta_{m_s} + \beta_{m_i})\Psi_{m_s, m_i} \\
                          & + i \gamma_{m_s, m_i} \exp(i \Delta\beta^{(0)} z) .
    \end{split}
\end{equation}
This allows us to change our representation to the real-space wave function, which is related to the modal formulation as $\Psi_{n_s, n_i} = \sum_{m_s, m_i}
\Psi_{m_s, m_i} \varepsilon_{n_s}^{(m_s)} \varepsilon_{n_i}^{(m_i)}$. Due to the orthogonality of the supermodes, we can invert these relations as $\Psi_{m_s, m_i} = \sum_{n_s, n_i} \Psi_{n_s, n_i} \varepsilon_{n_s}^{(m_s)} \varepsilon_{n_i}^{(m_i)}$. Substituting the latter expression in Eq.~\reqt{dPsi_m} and again employing the orthogonality property together with the fact that by definition the supermodes satisfy Eq.~\reqt{E1d}, yields the differential equations~\reqt{real_space_maintext} for the biphoton wave function in real-space representation.
\subsection{Experimental realization}
To realize the classical light simulator of quantum biphoton statistics, we fabricate waveguide arrays in fused silica using the femtosecond direct-write technique \cite{Itoh:2006bh, Szameit:2006-507:APB} where we used 160 fs pulses with an average power of 20 mW and a writing velocity of 60 mm/min. A sketch of the writing setup is shown in Fig.~\rpict{writingSetup}(a).
\pict{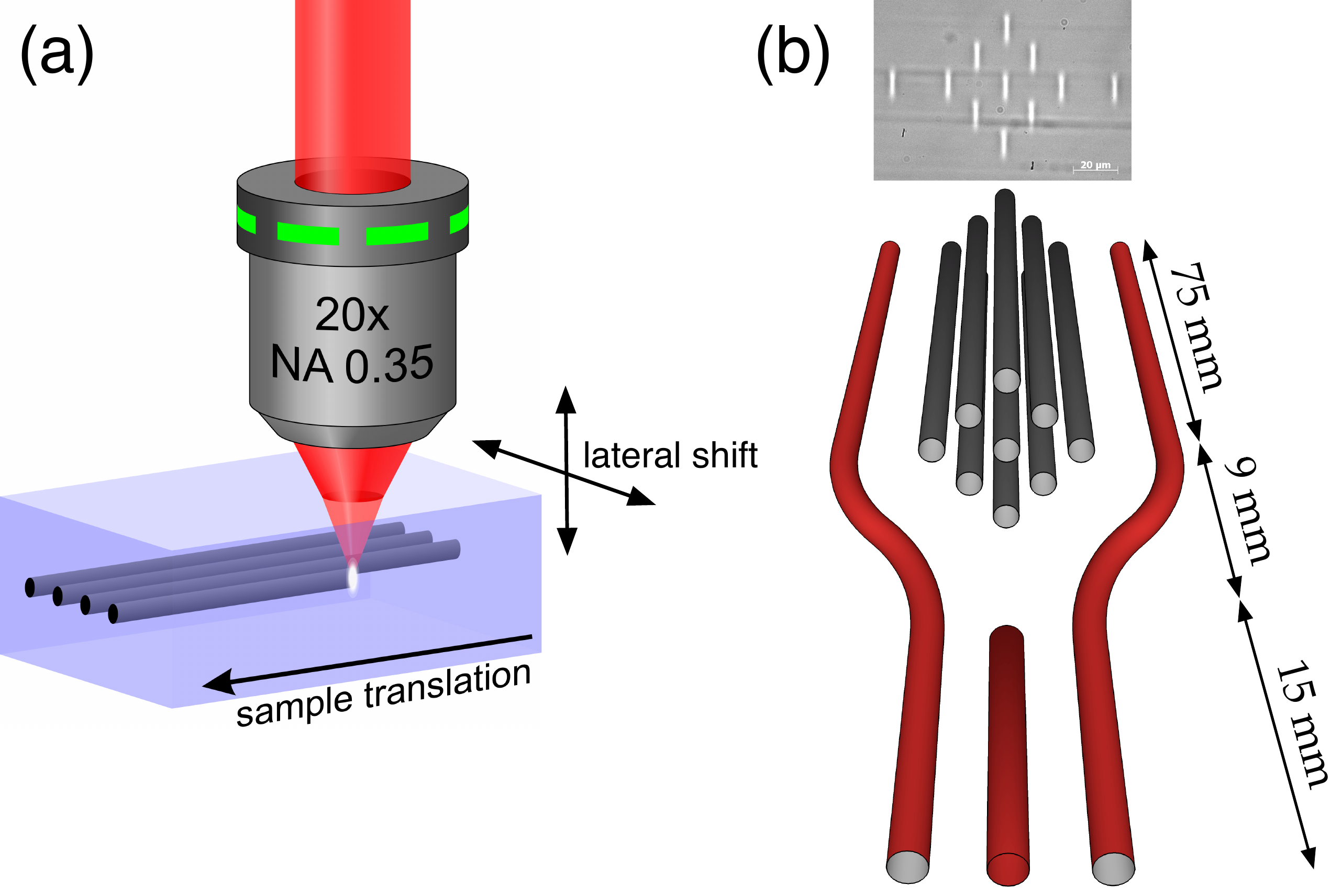}{writingSetup}{\textbf{Fabrication setting} \textbf{(a)}~Sketch of the writing setup, where femtosecond laser pulses are focused into a
transparent bulk material. In the focal region, the refractive index of the material is permanently increased. \textbf{(b)}~Sketch of the setting
with two edge pump waveguides, where the phase-matched symmetric amplitude distribution of the pump is achieved by a specially designed
three-waveguide directional coupler. The light is launched into the central waveguide marked with an arrow. Additionally, a micrograph of the output
facet is shown as inset (scale bar equals 40 $\mu$m).}
Our optical simulator consists of a $3 \times 3$ waveguide array with outer pump waveguides coupled to the corners of the array as shown in Fig.~\rpict{writingSetup}(b). To ensure homogenous horizontal and vertical coupling we rotated the structure by 45$^\circ$ and fabricated the waveguides in a rhombic geometry. To meet the undepleted pump approximation of SPDC we made sure that the intensities in the pump waveguide do not decrease significantly by choosing a sufficiently weak coupling of $C_{p}=0.125$ cm$^{-1}$, corresponding to a distance between the pump waveguide and the array of 25 $\mu$m~\cite{Szameit:2007cr}. Laser light of 633 nm wavelength was injected into the pump waveguides and the output intensity ($I_{n_s, n_i}$) was observed with a CCD camera. We then use the measured intensity to determine the simulated biphoton correlation.

In a first step we investigated a simulator having a single pump waveguide [Fig.~\rpict{fig1pump}(a)]. The distance between the array waveguides is 17 $\mu$m corresponding to a coupling constant of $C=0.78$~cm$^{-1}$.

The second step was an experiment performed with two pump waveguides at the array corners where we use a specially designed three-waveguide directional coupler that allows us to symmetrically distribute the light over the pump waveguides in a phase-matched fashion [Fig.~\rpict{fig2pump}(a), Fig.~\rpict{writingSetup}(b)].
Here we chose an inner distance of 21 $\mu$m which is best fitted by the numerical calculation with a coupling constant of $C=0.26$~cm$^{-1}$.
\section{References}


%
\section{Acknowledgements}
The authors wish to thank the German Ministry of Education and Research (Center for Innovation Competence program, grant 03Z1HN31). Robert Keil is supported by the Abbe School of Photonics Jena.
\section{Author contributions}
A.~S.~S., A.~A.~S. and Yu.~S.~K. suggested the concept and did the theoretical analysis; M.~G. and A.~S.~S. performed the numerical calculations; M.~G., A.~S.~S., R.~K. and A.~S. proposed the experimental realization; M.~G. fabricated the samples and performed the measurements; M.~G., A.~S.~S., R.~K., A.~A.~S., M.~H., A.~S. and Yu.~S.~K. discussed the results; and all authors cowrote the manuscript.

\end{document}